\documentclass{svmult}
\usepackage{natbib}
\usepackage{makeidx}
\usepackage{graphicx}
\usepackage{multicol}
\usepackage{url}

\newcommand{\enzo}{\code{Enzo}}
\newcommand{\code}[1]{\textsf{#1}}

\makeindex

\begin{document}
%
%
\title*{Introducing \enzo, an AMR Cosmology Application}
\author{Brian W. O'Shea\inst{1},
Greg Bryan\inst{2},  
James Bordner\inst{1}, 
Michael L. Norman\inst{1},
Tom Abel\inst{3},
Robert Harkness\inst{4} and Alexei Kritsuk\inst{1}}
\institute{Laboratory for Computational Astrophysics at the 
Center for Astrophysics and Space Sciences, 
University of California at San Diego, La Jolla, CA 92093, USA
\texttt{bwoshea@cosmos.ucsd.edu}
\and Department of Physics, University of Oxford, Oxford Ox1 3RH
\and Department of Astronomy and Astrophysics, Penn State University, University Park, PA 16802, USA
\and San Diego Supercomputing Center, UCSD, La Jolla, CA 92093, USA}
\titlerunning{Introducing \enzo, an AMR Cosmological Code}
\authorrunning{B.W. O'Shea et al.}
\maketitle

%
%
\textbf{Abstract.}  In this paper we introduce \enzo, a 3D
MPI-parallel Eulerian block-structured adaptive mesh refinement cosmology
code.  \enzo\ is designed to simulate cosmological structure formation,
but can also be used to simulate a wide range of astrophysical situations.
\enzo\ solves dark matter N-body dynamics using the 
particle-mesh technique.  The Poisson equation is
solved using a combination of fast fourier transform (on a periodic
root grid) and multigrid
techniques (on non-periodic subgrids).  Euler's equations
of hydrodynamics are solved using a modified version of the 
piecewise parabolic method.  Several additional physics packages are 
implemented in the code, including several varieties of radiative
cooling, a metagalactic ultraviolet background, and prescriptions 
for star formation and feedback.  We also show results illustrating 
properties of the adaptive mesh portion of the code.  Information
on profiling and optimizing the performance of the code can be found
in the contribution by James Bordner in this volume.

%
%
\section{Introduction}\label{sec:intro}

In astrophysics in general, and cosmology in particular, any given
object of interest can have many important length and time scales.  An excellent
example of this is the process of galaxy formation.  When studying the 
assembly of galaxies in a cosmological context, one wants to resolve
a large enough volume of the universe to capture enough large-scale 
structure (a box with length on the order of several 
megaparsecs\footnote{1 parsec = 3.26 light years = $3.0857 \times 10^{18} cm$}).  
However, in
order to adequately resolve structure in an individual galaxy one wants
to have resolution two orders of magnitude smaller than the ultimate
size of the objects of interest (a dwarf galaxy is on the order of $\sim$
1 kiloparsec).  This typical cosmological problem requires roughly
five orders of magnitude of dynamical range, which is prohibitively
expensive when done using a single grid.  Many other astrophysical phenomena,
such as the study of molecular clouds and the formation of stars and galaxy 
clusters, require similarly large dynamical range.  Many scientists have
adopted Lagrangean techniques such as smoothed particle hydrodynamics 
(SPH)\cite{Monaghan92}
to address these issues.  However, this type of method suffers from several
drawbacks, including poor shock resolution and fixed mass resolution in regions
of interest.  The use of grid-based techniques with structured adaptive 
mesh refinement avoids many of these problems, and additionally allows the use
of higher-order hydrodynamics schemes.

In this paper we present \enzo, an MPI-parallel 3D Eulerian 
adaptive mesh refinement code.
Though it was originally designed to study cosmological structure formation, 
the code is extensible and can be used for a wide range of astrophysical 
phenomena. For more information on the performance and optimization of the 
\enzo\ code, see the contribution by James Bordner in this volume.  The \enzo\ web page,
which contains documentation and the source code, can be found at
\url{http://cosmos.ucsd.edu/enzo/}.  

%
%
\section{Methodology}\label{sec:method}

\subsection{Application and AMR Implementation}
\enzo\ is developed and maintained by the 
Laboratory for Computational Astrophysics at the University of California in
San Diego.  The code is written in a mixture of C++ and Fortran 77.
High-level functions and 
data structures are implemented in C++ and computationally intensive
lower-level functions are implemented in Fortran.  \enzo\ is parallelized
using the MPI message-passing 
library\footnote{\url{http://www-unix.mcs.anl.gov/mpi/}} and uses the
HDF5\footnote{\url{http://hdf.ncsa.uiuc.edu/HDF5/}} data format
to write out data and restart files in a platform-independent format.
The code is quite
portable and has been ported to numerous parallel shared and distributed
memory systems, including the IBM SPs and p690 systems, SGI Origin 2000s and
numerous Linux Beowulf-style clusters.

The code allows hydrodynamic and N-body simulations in 1, 2 and 3 dimensions 
using the structured adaptive mesh refinement of 
Berger \& Colella\cite{Berger89}, and allows arbitrary integer ratios of parent
and child grid resolution and mesh refinement based on a variety of criteria,
including baryon and dark matter overdensity or slope, the existence of shocks, 
Jeans length, and cell cooling time.  The code can also have fixed static
nested subgrids, allowing higher initial resolution in a subvolume of the 
simulation.  Refinement can occur anywhere within the simulation volume or
in a user-specified subvolume.

The AMR grid patches are the primary data structure in \enzo.  Each individual 
patch is treated as an individual object, and can contain both field variables
and particle data.  Individual patches are organized into a dynamic distributed
AMR mesh hierarchy using arrays of linked lists to pointers to grid objects.
The code uses a simple dynamic load-balancing scheme to distribute the workload 
within each level of the AMR hierarchy evenly across all processors.

\begin{figure}
\centering
\includegraphics[height=5 cm]{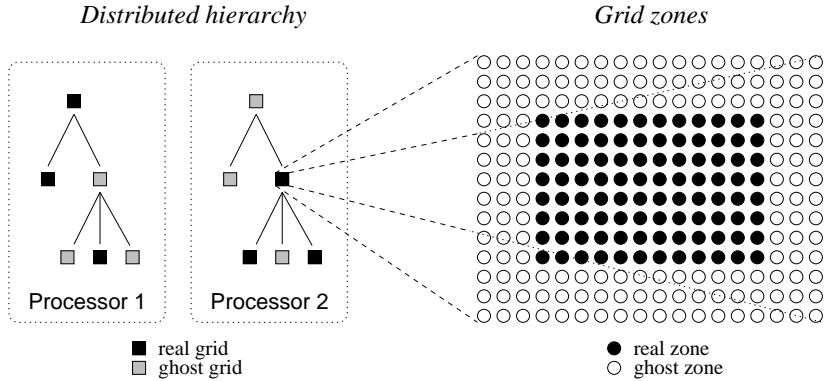}
\caption{Real and ghost grids in a hierarchy; real and ghost zones
in a grid.}
\label{fig.amrhier}
\end{figure}

Although each processor stores the entire distributed AMR hierarchy, not all
processors contain all grid data.  A grid is a \emph{real grid} on a 
particular processor if its data is allocated to that processor, and a 
\emph{ghost grid} if its data is allocated on a different processor.  Each
grid is a real grid on exactly one processor, and a ghost grid on all others.
When communication is necessary, MPI is used to transfer the mesh or particle
data between processors.  The tree structure of a small illustrative 2D AMR
hierachy -- six total grids in a three level hierarchy distributed across
two processors -- is shown on the left in Figure~\ref{fig.amrhier}.

Each data field on a real grid is an array of zones with dimensionality
equal to that of the simulation (typically 3D in cosmological structure
formation).  Zones are partitioned into a core block of \emph{real zones}
and a surrounding layer of \emph{ghost zones}.  Real zones are used to
store the data field values, and ghost zones are used to temporarily store
neighboring grid values when required for updating real zones.
The ghost zone layer is three zones deep in order to accomodate the computational
stencil in the hydrodynamics solver (Section~\ref{subsec:hydro}), as indicated
in the right panel in Figure~\ref{fig.amrhier}.  These ghost
zones can lead to significant computational and storage overhead, especially
for the smaller grid patches that are typically found in the deeper levels of an 
AMR grid hierarchy.

For more information on \enzo\ implementation and data structures, see 
references \cite{Bryan99}, \cite{Bryan97}, \cite{Bryan99a} and \cite{Norman99}.

\subsection{N-body Dynamics}\label{subsec:gravity}

The dynamics of large-scale structures are dominated by ``dark matter,'' 
which accounts for $\sim$ 85\% of the matter in the universe but can only
influence baryons via gravitational interaction. 
There are many other astrophysical situations where gravitational 
physics is important as well, such as galaxy collisions, where the stars in
the two galaxies tend to interact in a collisionless way.

There are multiple ways that one can go about calculating the gravitational 
potential (which is an elliptical equation in the Newtonian limit) in 
a structured AMR framework.  One way would be to model the dark matter (or other
collisionless particle-like objects, such as stars) 
as a second fluid in addition to the baryon fluid and solve the collisionless
Boltzmann equation, which follows the evolution of the fluid density in
both physical space and velocity space (referred to collectively as
``phase space''.  This is computationally prohibitive due to the large 
dimensionality of the problem and because the interesting portion of the 
solution to the equation does not tend to occupy a small volume of the
computational domain, which makes this approach unappealing in the 
context of an AMR code.

\enzo\ uses a totally different approach to collisionless systems, namely,
the N-body method.  This method follows trajectories of a representative 
sample of individual particles and is much more efficient than a direct 
solution of the Boltzmann equation in most astrophysical situations.
The particle trajectories are controlled by a simple set of coupled equations
(for simplicity, we omit cosmological terms):

\begin{equation}
\frac{d\vec{x_p}}{dt} = \vec{v_p}
\end{equation}

\begin{equation}
\frac{d\vec{v}_p}{dt} = -\nabla\phi
\end{equation}

Where $\vec{x_p}$ and $\vec{v_p}$ are the particle position and
velocity vectors, respectively, and the term on the right-hand side of 
the second equation is the gravitational force term.  The solution to this
can be found by solving the elliptic Poisson's equation:

\begin{equation}
\nabla^2 \phi = 4\pi G \rho
\end{equation}

where $\rho$ is the density of both the collisional fluid (baryon gas)
and the collisionless fluid (particles).

These equations are finite-differenced and for simplicity are 
solved with the same timestep as the equations of hydrodynamics.
The dark matter particles are sampled onto the grids using the 
triangular-shaped cloud (TSC) interpolation technique to form a
spatially discretized density field (analogous to the baryon
densities used to calculate the equations of hydrodynamics)
and the elliptical equation is solved using FFTs on the
triply periodic root grid and multigrid relaxation on the subgrids.
Once the forces have been computed on the mesh, they are interpolated
to the particle positions where they are used to update their 
velocities.

\subsection{Hydrodynamics}\label{subsec:hydro}

The primary hydrodynamic method used in \enzo\ is based on the 
piecewise parabolic method (PPM) of Woodward \& Colella~\cite{Woodward84}
which has been significantly modified for the study of cosmology.  The
modifications and several tests are described in Bryan et al.~\cite{Bryan95},
but we provide a short description here.

PPM is a higher-order-accurate version of Godunov's method with 
third-order-accurate piecewise parabolic monotolic interpolation 
and a nonlinear Riemann solver for shock capturing.  It does an excellent
job capturing strong shocks and outflows.  Multidimensional schemes are 
built up by directional splitting, and produce a method that is formally 
second-order-accurate in space and time and explicitly conserves energy, 
momentum and mass flux.  The conservation laws for fluid mass, momentum
and energy density are written in comoving coordinates for a
Friedman-Robertson-Walker spacetime.  Both the conservation laws and
Riemann solver are modified to include gravity, as calculated in
Section~\ref{subsec:gravity}.

There are many situations in astrophysics, such as the bulk hypersonic
motion of gas, where the kinetic energy of a fluid can dominate its internal 
energy by many orders of magnitude.  In these situations, limitations on
machine precision can cause significant inaccuracy in the calculation of 
pressures and temperatures in the baryon gas.  In order to address this 
issues, \enzo\ solves both the internal gas energy equation and the total 
energy equation everywhere on each grid, at all times.  This 
\emph{dual energy formalism} ensures that the method yields the correct
entropy jump at strong shocks and also yields accurate pressures and 
temperatures in cosmological hypersonic flows.

As a check on our primary hydrodynamic method, we also include an
implementation of the hydro algorithm used in the Zeus astrophysical
code.  \cite{Stone92a, Stone92b}  This staggered grid, finite difference
method uses artificial viscosity
as a shock-capturing technique and is formally first-order-accurate when
using variable timesteps (as is common in structure formation simulations), 
and is not the preferred method in the \enzo\ code.

\subsection{Additional Physics Packages}\label{subsec:morephysics}

Several physics packages are implemented in addition to dark matter and
adiabatic gas dynamics.  The cooling and heating of gas is extremely
important in astrophysical situations.  To this extent, 
two radiative cooling models and several uniform ultraviolet background 
models have been implemented in an easily extensible framework.

The simpler of the two radiative cooling models assumes that all
species in the baryonic gas are in equilibrium and calculates cooling 
rates directly from a cooling curve assuming $Z = 0.3~Z_{\odot}$.  The 
second routine, developed by Abel, Zhang, Anninos \& Norman~\cite{Abel97,Anninos97},
assumes that the gas has primordial abundances (ie, a gas which is composed of
hydrogen and helium, and unpolluted by metals), and solves
a reaction network of 28 equations which includes collisional and radiative processes
for 9 seperate species ($H, H^+, $He, He$^+$, He$^{++}, H^-, H_2^+, H_2,$ and $e^-$.  In 
order to increase the speed of the calculation, this method takes the reactions with 
the shortest time scales (those involving $H^-$ and $H_2^+$) and decouples them from
the rest of the reaction network and imposes equilibrium concentrations, which is 
highly accurate for cosmological processes.  See Anninos et al.~\cite{Anninos97} and 
Abel et al.~\cite{Abel97} for more information.

The vast majority of the volume of the present-day universe is occupied by
low-density gas which has been ionized by ultraviolet radiation from quasars, 
stars and other sources.  This low density gas, collectively referred to as the
``Lyman-$\alpha$ Forest'' because it is primarily observed as a dense collection of
absorption lines in spectra from distant quasars (highly luminous extragalactic
objects), is useful because it can be used to determine several cosmological
parameters and also as a tool for studying the formation and evolution of 
structure in the universe (see \cite{Rauch98} for more information).  The 
spectrum of the ultraviolet radiation background plays an important part
in determining the ionization properties of the Lyman-$\alpha$ forest, so 
it is very important to model this correctly.  To this end, we have
implemented several models for uniform ultraviolet background radiation
based upon the models of Haardt \& Madau~\cite{haardt96}.

One of the most important processes when studying the formation and
evolution of galaxies (and to a lesser extent, groups and clusters of 
galaxies and the gas surrounding them) is the formation and feedback
of stars.  We use a heuristic prescription similar to that of 
Cen \& Ostriker~\cite{Cen92} to convert gas which is rapidly
cooling and increasing in density into star ``particles'' which
represent an ensemble of stars.  These particles then evolve 
collisionlessly while returning metals and thermal energy back
into the gas in which they formed via hot, metal-enriched winds.  

As mentioned in Section~\ref{sec:intro}, \enzo\ can be downloaded
from the web at \url{http://cosmos.ucsd.edu/enzo/}.  Vigorous code 
development is taking place, and we are in the process of adding
ideal magnetohydrodynamics and a flux-limited radiation diffusion
scheme to our AMR code, which will significantly enhance the 
capabilities of the code as a general-purpose astrophysical tool.

%
%
\section{Adaptive Mesh Characteristics}\label{sec:results}

\begin{figure}
\centering
\includegraphics[height=8 cm]{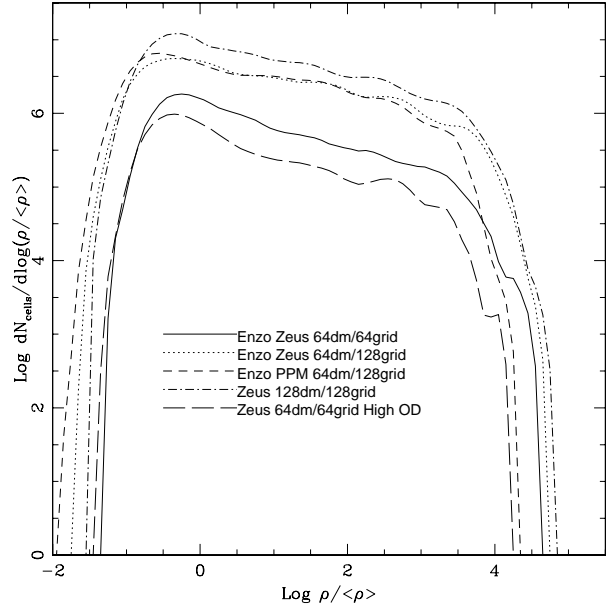}
\caption{Number of cells as a function of baryon overdensity
(normalized by bin size) for a representative suite of \enzo\
cosmology simulations.   The simulations are labelled such that 
the first number corresponds to the number of dark matter particles
and the second number shows the root grid size, ie, 64dm/128grid means 
that the simulation has $64^3$ dark matter particles and a $128^3$ root 
grid.  Simulations with two different hydrodynamic methods are used.
PPM:  Piecewise Parabolic Method.~\cite{Bryan95}  Zeus:  The
hydro method  used in the Zeus astrophysical code.~\cite{Stone92a} }
\label{fig.cellod}
\end{figure}

The adaptive nature of grid cells in the AMR simulations
results in a wide range of baryon mass scales being resolved.
Figure~\ref{fig.cellod} shows the distribution of cells
as a function of overdensity for a range of \enzo\ simulations 
in a simulation volume which is 3 $h^{-1}$ megaparsecs
 on a side.  These simulations use 
either a $64^3$ or $128^3$ root grid and either 5 or 6 
levels of refinement (such that $L_{box}/e = 4096$, where
$L_{box}$ is the box size and $e$ is the smallest spatial
scale that can be resolved).
All grids are refined by a factor
of 2.0, and grids are refined when dark matter density (baryon density) 
exceeds a factor of 4.0 (2.0) times the mean density
of cells at that level.  In addition, a 
simulation is performed where the overdensity
threshold is doubled.  
Initial conditions are generated using power spectra and
methods common to cosmological simulations.
Examination of Figure~\ref{fig.cellod} 
shows that the entire density range in the simulations is
covered by large numbers of cells.
In particular, cells at low densities are well-resolved in these
simulations, which is in stark contrast to simulations performed
using Lagrangian methods, which are typically undersampled at
low density.  Raising the overdensity threshold for refinement 
decreases the total number of cells but their relative distribution
as a function of overdensity is 
unchanged.  In all simulations the total number of cells at
the end of the run has increase by a factor of $\sim 8-10$
from the number of cells in the root grid.

\begin{figure}
\centering
\includegraphics[height=7 cm]{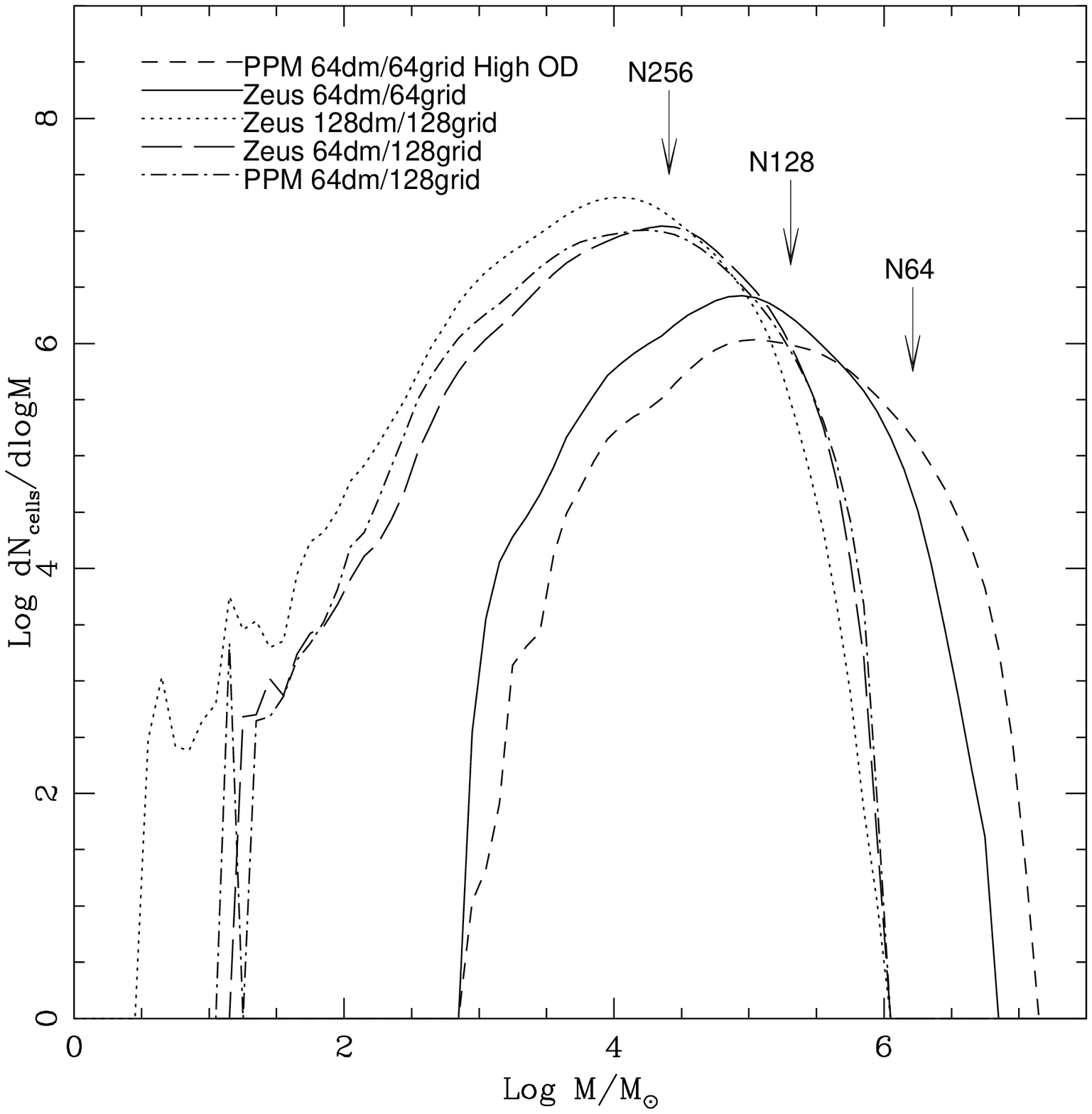}
\caption{Number of cells as a function of baryonic mass (normalized to bin size)
for several \enzo\ simulations with $L_{box} = 3 h^{-1} Mpc$.  The arrows
correspond to the mean mass resolution of the root grid for simulation volumes with
the same volume but root grids with $64^3, 128^3$ or $256^3$ cells (labelled N64, N128 
and N256, respectively).  See Figure~\ref{fig.cellod} for a description of 
the line labels.}
\label{fig.cellmass}
\end{figure}

Figure~\ref{fig.cellmass} shows the distribution of number of cells
at the end of a simulation as a function of the mass of baryons in
that cell.  Arrows indicate the mean cell mass contained on the
root grid at the onset of the simulation for simulations covering
the same spatial volume as the simulations described above with a
$64^3, 128^3$ or $256^3$ root grid (labelled N64, N128 and N256,
respectively).  Over the course of the simulation the mean mass
resolution, as indicated by the peak of the distribution, increases
by almost an order of magnitude relative to the initial mass resolution,
though the distribution of cell masses is quite large.  Figure~\ref{fig.cellmass}
shows that the mean cell mass as a function of overdensity (at the end of
the simulation run) stays fairly constant, which lower mean cell masses in
underdense regions and higher mean cell masses in highly overdense regions
(presumably due to the limitation on the number of levels of adaptive mesh 
refinement allowed).  The mean cell mass over the entire density range is
between $\sim 5-10$ times better than the starting mass resolution for all
simulations.  Runs with lower overdensity criteria for refinement
have somewhat better mass resolution overall.

\begin{figure}
\centering
\includegraphics[height=7 cm]{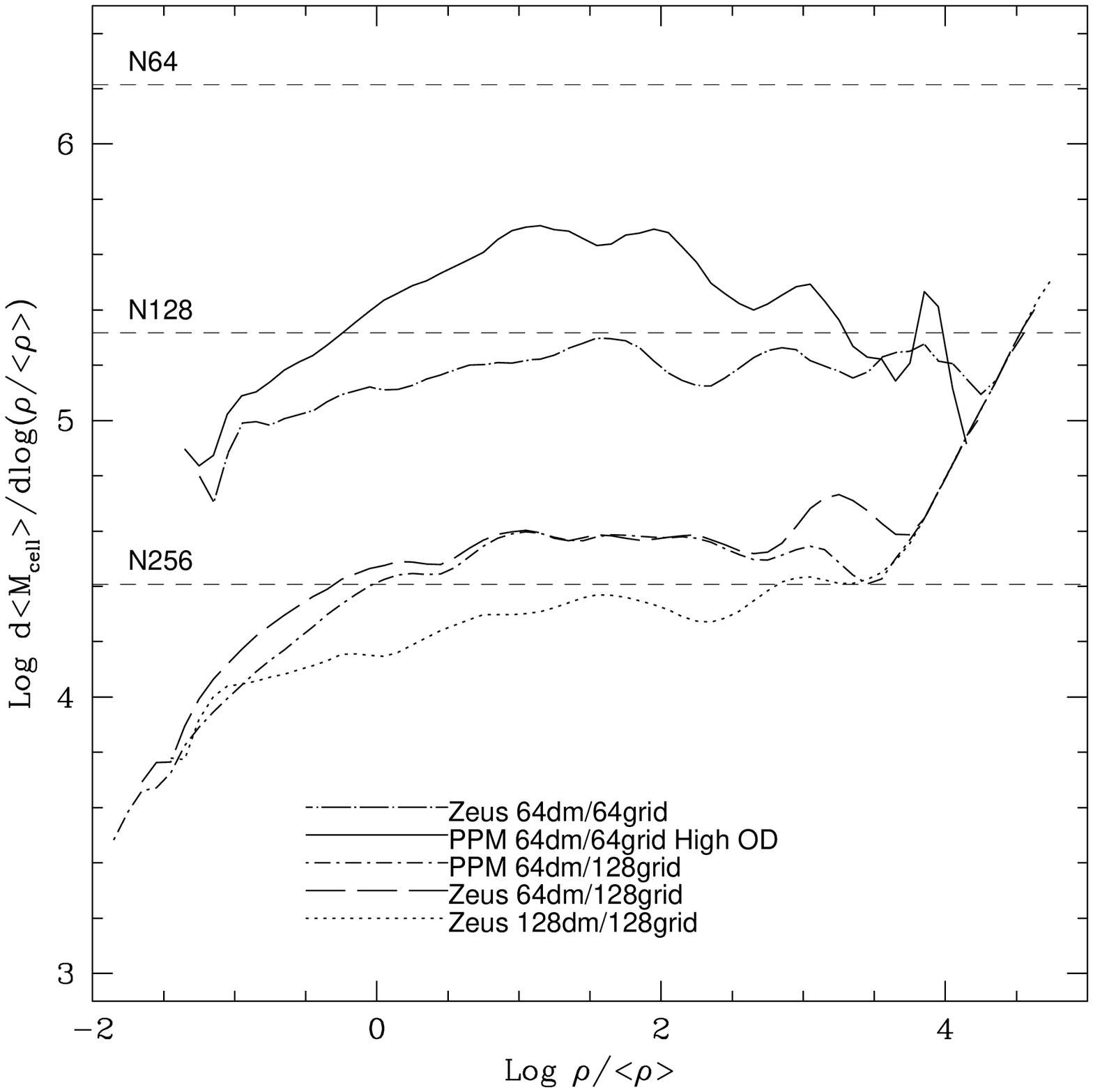}
\caption{Mean baryonic mass in cells as a function of baryon overdensity (normalized to
bin size) in \enzo\ simulations.  These are the same simulations (with the same labels) as
in Figure~\ref{fig.cellod}.  Horizontal lines correspond to the initial mean mass resolution
of simulations with the same volume but $64^3, 128^3$ and $256^3$ root grid cells (labelled N64,
N128 and N256, respectively).}
\label{fig.cellmassod}
\end{figure}


%
%
\section{Summary}\label{sec:summary}

In this paper we have presented \enzo\, a cosmology code
which combines collisionless N-body particle dynamics with 
a hydrodynamics package based on the piecewise parabolic 
method, all within a block-based adaptive mesh refinement
algorithm.  Several other physics packages are implemented,
including multiple models for gas cooling and ionization, a
uniform ultraviolet background model for gas heating, and a 
prescription for star formation and feedback.

\enzo\ is being released to the public as a community 
astrophysical simulation code.  This code is being
modified and documented to be as widely useful as possible,
and can be found at \url{http://cosmos.ucsd.edu/enzo/}.  Active
development is taking place, centering on the addition of 
magnetohydrodynamics and a diffusive radiative transfer
algorithm.

Further information concerning the performance of the \enzo\ code
(including a package of performance monitoring and visualization
tools) is described in a contribution by James Bordner in this volume, 
and on the \enzo\ website.

\newpage

%
%
\bibliographystyle{unsrt}


\printindex
\end{document}